# X-ray photoemission study of polycrystalline $Mg_{1-x}Al_xB_2$

Fa-Min Liu[1,2], C. Dong[1], J. Q. Li[1], T. M. Wang[2]

[1] National Laboratory for Superconductivity, Institute of Physics, Chinese Academy of Sciences, Beijing 100080, P. R. China

[2] Center of Material Physics and Chemistry, Beijing University of Aeronautics & Astronautics, Beijing 100083, P. R. China

**Abstract:** Polycrystalline $Mg_{1-x}Al_xB_2$ (x = 0, 0.1, 0.4, 0.5, 0.6, 1) samples have been prepared by a conventional solid state reaction. The Mg 2p, Al 2p and B 1s core level binding energies have been studied using X-ray photoemission spectroscopy (XPS). It is found that Mg 2p, Al 2p and B 1s core level binding energies increase with the Al content in $Mg_{1-x}Al_xB_2$ when $x \leq 0.5$, and the maxima in the binding energies of Mg 2p and Al 2p are observed in sample $Mg_{0.5}Al_{0.5}B_2$. In contrast to the Mg 2p and Al 2p core levels, the B 1s binding energy shows a minimum when x = 0.5.

Corresponding Author:

Fa-Min Liu
Group CSC02
National Laboratory for Superconductivity
Institute of Physics, Chinese Academy of Sciences,
Beijing 100080, P. R. China
E-mail: fmliu@aphy.iphy.ac.cn
Tel. 86-10-82649180,   Fax. 86-10-82649531

1 Introduction

The discovery of superconductivity in the metallic compound $MgB_2$[1] with its transition temperature at 39K has stimulated considerable interest in the area of superconductivity. Intensive studies have been carried out on this new material by means of magnetic,[2] microstructure,[3] Raman and Infrared spectra,[4,5] transport,[6,7] Hall measurements [8] and other experimental techniques.[9,10] Doping Al on the Mg site can introduce electrons into the bands and lead to the loss of superconductivity as observed in $Mg_{1-x}Al_xB_2$ materials[11] due to the a structural instablity for x > 0.1. This negative doping effect on $T_C$ could be explained by a strong electron-phonon coupling mechanisms in BCS model. Lorenz, et al. [12] have studied the thermoelectric power S and resistivity ρ of $Mg_{1-x}Al_xB_2$. They show that S is positive and increases linearly with temperature above the superconducting transition temperature $T_C$, and also, $T_C$ and $T_0$ both decrease with Al doping whereas the slope of S(T) in the linear range increase with the Al content. Recently we have reported a rich variety of phenomena resulting from Al ordering along the c-axis direction.[13] Then a strong tendency for the formation of a superstructure has been studied from systematic ab initio calculation.[14] X-ray photoemission spectroscopy (XPS) is a powerful tool for characterizing surfaces and has been used to study effects on YBCO superconductor for various oxygen environments and heat treatments.[15,16] In this article we mainly report the X-ray photoemission spectra in $Mg_{1-x}Al_xB_2$. Maxima in the binding energies of Mg/Al 2p and a minimum of B 1s core level spectra are observed as a signature for a stable superlattice structure of $Mg_{0.5}Al_{0.5}B_2$.

## 2 Experimental

Polycrystalline samples of $Mg_{1-x}Al_xB_2$ were prepared by a conventional solid state reaction. The $Mg_{1-x}Al_xB_2$ sample used in XPS experiments was prepared from Mg (99.5%) powder of 50 mesh, B (99.999%) powder of 325 mesh and Al (99.0%) powder of 200 mesh. Stoichiometric mixture of the starting materials Mg, Al and B powder were pressed into pellets, wrapped in a Ta foil and sealed in a quartz ampoule. The samples were sintered at 900 $^o$C for 2 hours and then cooled down to room temperature. The phase structures were analyzed by an M18AHF type X-ray diffractometer (XRD) using Cu-K$\alpha$ radiation. The microstructure of the samples was studied using a H-9000NA transmission electron microscopy (TEM) with an atomic resolution of about 0.19 nm. Specimens for transmission-electron microscopy observations were polished mechanically with a Gatan polisher to a thickness of around 50μm and then ion-milled by a Gatan-691 PIPS ion miller for 3 h. Electrical resistance and the superconducting transition were investigated using the standard four probe method.

X-ray photoemission spectra data were obtained with a VG ESCALAB MK II spectrometer using Mg K$_\alpha$ source radiation operating at an accelerating voltage of 12 kV and an emission current of 20 mA. The working pressure in the XPS chamber was approximately $6 \times 10^{-7}$ Pa. Survey spectra were collected with a pass energy of 50 eV. Spectra are measured at room temperature with photoemission 75$^o$ from the surface normal for the polycrystalline $Mg_{1-x}Al_xB_2$ pellets. Specimens for XPS observations were ground on the abrasive paper (CW2000), and then were finished with finishing machinery.

3 Results and discussion

The results of XRD, TEM, Raman spectra, and superconductor properties for the $Mg_{1-x}Al_xB_2$ materials had been reported elsewhere.[13] In this paper, we mainly present the X-ray photoemission spectra data of $Mg_{1-x}Al_xB_2$ samples, and discuss the changes of the Mg 2p, Al 2p and B 1s core levels with the Al content variation.

Fig.1 and Table1 show the Mg 2p core level of polycrystalline $Mg_{1-x}Al_xB_2$. One can see that the Mg 2p core level are at about 50.4, 50.6, 50.7, 51.3, and 50.8 eV, corresponding to x = 0, 0.1, 0.4, 0.5, and 0.6, respectively. Compared with that of Mg metal, the Mg 2p (49.8 eV) core level [17] of the $Mg_{1-x}Al_xB_2$ samples have a large chemical shifts of 0.6, 0.8, 0.9, 1.5, 1.0 eV, corresponding to x = 0, 0.1, 0.4, 0.5, and 0.6, respectively. These data are different from that in pure $MgB_2$ film (49.4 eV) reported by Vasquez, et al.[18] and Ueda, et al..[19] From Fig.1, the Mg 2p core levels of $Mg_{1-x}Al_xB_2$ with $0 \leq x \leq 0.1$ and $0.25 \leq x \leq 0.4$ have small changes because all these samples were found to be single-phased ($AlB_2$ type) by powder X-ray diffraction.[11] This crystal structure consists of honeycomb-net planes of boron, separated by triangular planes of the metals. In polycrystalline $Mg_{0.5}Al_{0.5}B_2$, the Mg2p core level is much higher than that of $MgB_2$ and MgO.[18] This signifies that polycrystalline $Mg_{0.5}Al_{0.5}B_2$ has a new structure. The Mg 2p core levels of $Mg_{1-x}Al_xB_2$ with $0.6 \leq x \leq 1$ also have small changes because their structures were all found to be $AlB_2$ type.[13]

Fig. 2 and Table 1 show the Al 2p core level of $Mg_{1-x}Al_xB_2$. One can see that the peak of Al 2p core levels of $Mg_{1-x}Al_xB_2$ are at about 73.9, 74.1, 75.8, 73.8, and 74.5 eV, for samples with x = 0.1, 0.4, 0.5, 0.6, and 1, respectively. These data are larger than that of Al metal (72.9 eV),[17] and they are close to the Al 2p core level (74.7 eV) of $Al_2O_3$. It is obvious that the Al 2p core levels of $Mg_{1-x}Al_xB_2$ samples have a chemical shifts of 1.0, 1.2, 2.9, 0.9, and 1.6 eV with respect to that of Al metal, corresponding to x = 0.1, 0.4, 0.5, 0.6, and 1, respectively. From Fig.2, one can note that $Mg_{1-x}Al_xB_2$ with $0 \leq x \leq 0.40$ and $0.6 \leq x < 1$ have approximately same Al 2p core level. However, it is noted that the Al 2p core level (75.8 eV) in $Mg_{0.5}Al_{0.5}B_2$ sample is close to that of $LiAlH_4$ (75.6 eV).[17] This new superstructure phase has been observed by TEM recently,[13] and this superstructure has an ideal composition of $MgAlB_4$.

As described above, the change trend of Mg 2p and Al 2p core levels of $Mg_{1-x}Al_xB_2$ are as follows: For $Mg_{1-x}Al_xB_2$ samples with x from 0.1 to 0.5, the Mg 2p and the Al 2p core levels increase slightly with x, and reach a maximum in $Mg_{0.5}Al_{0.5}B_2$ sample. Then, the Mg 2p and the Al 2p core levels decrease slightly when x increases in $Mg_{1-x}Al_xB_2$ samples with $x \geq 0.6$. In $Mg_{1-x}Al_xB_2$ with x from 0.1 to 0.4, Mg was partially substituted by Al, and, Mg and Al atoms are randomly distributed. The $Mg_{0.5}Al_{0.5}B_2$ is a fully layered superstructure in which the Al and Mg layers are arranged alternately along the c-axis. Therefore, the Mg 2p and the Al 2p core levels come to maximum. For $Mg_{1-x}Al_xB_2$ samples with $0.6 \leq x \leq 0.9$, the Mg 2p and the Al 2p core levels restore to close to that of $Mg_{1-x}Al_xB_2$ with $0.1 \leq x \leq 0.4$ because the Al-rich samples exist two phases or the disordered structure again.

The XPS core level binding energies of B 1s of polycrystalline $Mg_{1-x}Al_xB_2$ are shown in Fig. 3 and Table 1. One can see that the B 1s core level consist of two peaks. The first peak is the core level of intermetallic $Mg_{1-x}Al_xB_2$, and the second peak is due to the boron oxides. Slusky, et al.[11] have shown that the two phase mixtures of two different $AlB_2$ –type phase will be observed in polycrystalline $Mg_{1-x}Al_xB_2$ with $0.1 \leq x < 0.25$. So in this region, $Mg_{1-x}Al_xB_2$ exists as two phases with significantly different c lattice parameters. The phase separation can also be observed in $Mg_{1-x}Al_xB_2$ with x near 0.7 to 0.75. From Fig. 3 and Table 1, the B 1s core level binding energies is somewhat close to that in transition-metal diborides, where the B1s binding energies are in the range 187.2-188.5 eV, [17,18,20] because $MgB_2$, $AlB_2$ and transition metal diborides belong to the common $AlB_2$ structural type. For samples with x = 0.6, the predominant phase also belongs to the $AlB_2$ structure type although it is in the two phase region. The B 1s signals of boron oxides are very strong in $Mg_{1-x}Al_xB_2$ with x = 0.1 and 0.4, indicating that the surfaces of these samples are oxidized extensively. $Mg_{0.5}Al_{0.5}B_2$ has a extraordinarily low B 1s core level (185.5 eV). This indicates that the B-B bonding of $Mg_{0.5}Al_{0.5}B_2$ differ from that of $MgB_2$. For $MgB_2$,[21] borons form a primitive honeycomb lattice, consisting of graphite-type sheets stacked with no displacement. However, in polycrystalline $Mg_{0.5}Al_{0.5}B_2$, the superstructure of $MgAlB_4$ was formed with an ordered Al and Mg layers, and boron layer were stacked with a small displacements. So polycrystalline $Mg_{0.5}Al_{0.5}B_2$ has the smallest B 1s core level binding energy.

Obviously, from Fig.3 and Table 1, one can noted that the B 1s core level of $Mg_{1-x}Al_xB_2$ has a special trend. For $0.1 \leq x \leq 0.4$, the B 1s core level was increase slightly with the Al content in $Mg_{1-x}Al_xB_2$ although the signal strength of B 1s is very weak. For $Mg_{0.5}Al_{0.5}B_2$, the B1s core level come to the minimum. And for $x \geq 0.6$, the B1s core level increases slightly again. We think that this phenomena can be interpreted by the bonding variation of Mg-B-Al. Al ions are in different Mg layer if $Mg_{1-x}Al_xB_2$ compound with x from 0.1 to 0.4. After ionic relaxation, the B ions shift from their original position towards the neighboring Al atoms. This leads to increase the B1s core level binding energies. In $Mg_{0.5}Al_{0.5}B_2$, Mg layer, Al layer and B layer is arranged in order. The arrangement of Mg-B-Al-B is very stable so that the B1s core level has a minimum. When x ≥ 0.6, Al-rich layer lead to change the arrangement of B-B site and increase of B1s core level. In addition, it is noted that the extent of $Mg_{1-x}Al_xB_2$ has been oxidized differently. For x = 0.1, 0.4 and 0.6, the B1s signal strength of boron oxides is larger than that of $Mg_{1-x}Al_xB_2$ owing to their disordered structure. In contrast, $Mg_{0.5}Al_{0.5}B_2$ has a stable superstructure. Their B1s signal strength is larger than that of boron oxides. The B1s core level of $Mg_{1-x}Al_xB_2$ has been affected by a mount of boron oxides.

As discussed above, one can note that the Mg 2p, Al 2p and B 1s core level binding energies of $Mg_{0.5}Al_{0.5}B_2$ differ from that of MgO, $Al_2O_3$, $MgB_2$ pellet and $MgB_2$ film. In polycrystalline $Mg_{0.5}Al_{0.5}B_2$, the Mg 2p core level (51.3 eV) is higher than that of Mg metal and $MgB_2$, and the Al 2p core level (75.8 eV) is much higher than that of Al metal and $Al_2O_3$, whereas, the B1s (185.5 eV) is much lower than that of B metal, $B_2O_3$, and $MgB_2$. It has shown that $Mg_{0.5}Al_{0.5}B_2$ has a new superlattice structure, and the superconducting transition temperature ($T_c$) of $Mg_{0.5}Al_{0.5}B_2$ decreases to 12 K.[13] Therefore, the B-B bonding plays an important role in the superconductivity of $MgB_2$ and related compounds.

4 Conclusions

In conclusion, we have studied X-ray photoemission spectra of $Mg_{1-x}Al_xB_2$. The Mg 2p and the Al 2p core level binding energies were increased slightly with Al content in $Mg_{1-x}Al_xB_2$, and the maximum binding of Mg 2p and Al 2p are observed in the sample of $Mg_{0.5}Al_{0.5}B_2$. In contrast, the B 1s core level binding energy has a minimum value in the $Mg_{0.5}Al_{0.5}B_2$ sample. These results show that $Mg_{0.5}Al_{0.5}B_2$ has a remarkable surface structural feature with special core level binding energies.

Acknowledgement:

This work reported here was supported by "Hundreds of Talents" program organized by the Chinese Academy of Sciences, P. R. China.

Caption for Figures and Table

Table 1 XPS core level binding energies of Mg2p, Al2p and B1s of polycrystalline $Mg_{1-x}Al_xB_2$

Fig.1 The Mg 2p core level of $Mg_{1-x}Al_xB_2$

Fig.2 The Al 2p core level of $Mg_{1-x}Al_xB_2$

Fig. 3 The B1s core level of $Mg_{1-x}Al_xB_2$

Table 1  XPS core level binding energies of Mg2p, Al2p and B1s of polycrystalline $Mg_{1-x}Al_xB_2$

| $Mg_{1-x}Al_xB_2$ | XPS core level binding energy ( eV ) | | | |
| --- | --- | --- | --- | --- |
| X | Mg2p | Al2p | B1s | |
| | | | $Mg_{1-x}Al_xB_2$ | Boron oxidized |
| 0 | 50.4 | | 186.6 | 192.1 |
| 0.1 | 50.6 | 73.9 | 187.8 | 192.4 |
| 0.4 | 50.7 | 74.1 | 188.0 | 192.5 |
| 0.5 | 51.3 | 75.8 | 185.5 | 191.1 |
| 0.6 | 50.8 | 73.8 | 186.4 | 191.8 |
| 1 | | 74.5 | 187.0 | 191.7 |

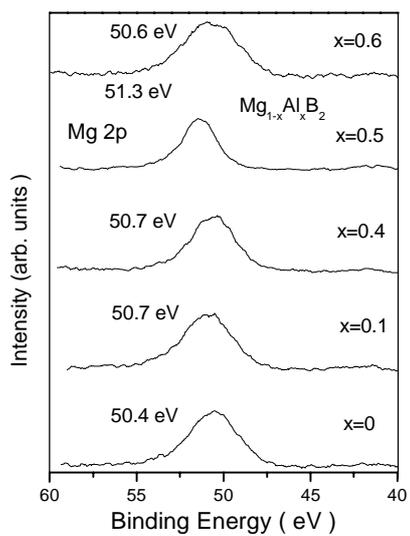

Fig.1 The Mg 2p core level of $Mg_{1-x}Al_xB_2$

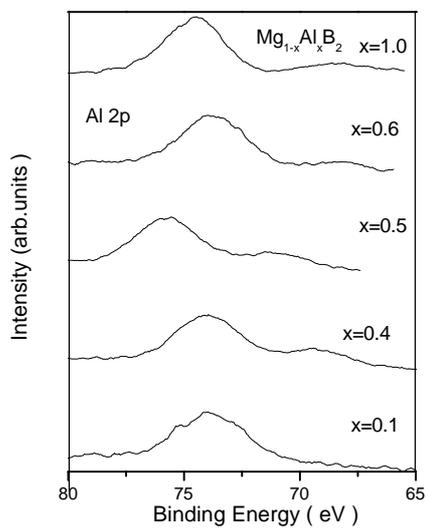

Fig.2 The Al 2p core level of $Mg_{1-x}Al_xB_2$

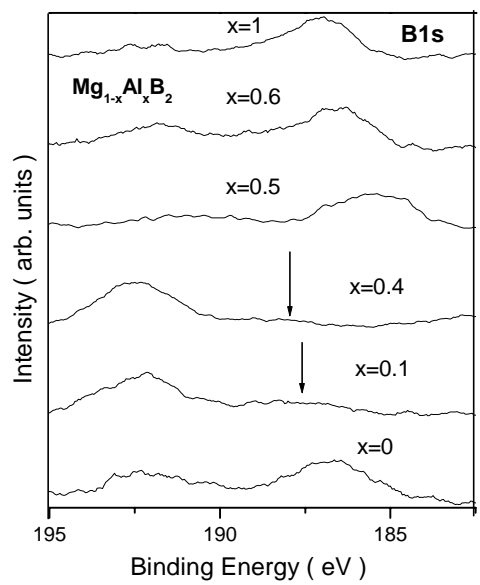

Fig. 3 The B1s core level of $Mg_{1-x}Al_xB_2$